\documentclass[floatfix,aps,prx,twocolumn]{revtex4-2}

\usepackage[T1]{fontenc}
\usepackage{lipsum}  
\usepackage{graphicx}
\usepackage[english]{babel}
\usepackage{upgreek}
\usepackage{amsmath}
\usepackage{amssymb}
\usepackage{tabls}
\usepackage{dsfont}
\usepackage[colorlinks=true,citecolor=blue,linkcolor=red,urlcolor=blue]{hyperref}

\makeatletter
\def\bbl@set@language#1{%
	\edef\languagename{%
		\ifnum\escapechar=\expandafter`\string#1\@empty
		\else\string#1\@empty\fi}%
	\@ifundefined{babel@language@alias@\languagename}{}{%
		\edef\languagename{\@nameuse{babel@language@alias@\languagename}}%
	}%
	\select@language{\languagename}%
	\expandafter\ifx\csname date\languagename\endcsname\relax\else
	\if@filesw
	\protected@write\@auxout{}{\string\select@language{\languagename}}%
	\bbl@for\bbl@tempa\BabelContentsFiles{%
		\addtocontents{\bbl@tempa}{\xstring\select@language{\languagename}}}%
	\bbl@usehooks{write}{}%
	\fi
	\fi}
\newcommand{\DeclareLanguageAlias}[2]{%
	\global\@namedef{babel@language@alias@#1}{#2}%
}
\makeatother

\newcommand\varpm{\mathbin{\vcenter{\hbox{%
  \oalign{\hfil$\scriptstyle+$\hfil\cr
          \noalign{\kern-.3ex}
          $\scriptscriptstyle({-})$\cr}%
}}}}
\newcommand\varmp{\mathbin{\vcenter{\hbox{%
  \oalign{$\scriptstyle({+})$\cr
          \noalign{\kern-.3ex}
          \hfil$\scriptscriptstyle-$\hfil\cr}%
}}}}

\DeclareLanguageAlias{en}{english}

\begin{document}

\title{Crosstalk analysis for simultaneously driven two-qubit gates in spin qubit arrays}

\author{Irina Heinz}
\email{irina.heinz@uni-konstanz.de}
\affiliation{Department of Physics, University of Konstanz, D-78457 Konstanz, Germany}
\author{Guido Burkard}
\email{guido.burkard@uni-konstanz.de}
\affiliation{Department of Physics, University of Konstanz, D-78457 Konstanz, Germany}


\begin{abstract}
	One of the challenges when scaling up semiconductor-based quantum processors consists in the presence of crosstalk errors caused by control operations on neighboring qubits.  In previous work, crosstalk in spin qubit arrays has been investigated for non-driven single qubits near individually driven quantum gates and for two simultaneously driven single-qubit gates. Nevertheless, simultaneous gates are not restricted to single-qubit operations but also include frequently used two-qubit gates such as the CNOT gate. We analyse the impact of crosstalk drives on qubit operations, such as the CNOT and CPHASE gates. We investigate the case of parallel $Y$ and CNOT gates, and we also consider a two-dimensional arrangement of two parallel CNOT gates and find unavoidable crosstalk. To minimize crosstalk errors, we develop appropriate control protocols.
\end{abstract}


\maketitle

\section{Introduction}
In the field of quantum computing several realization platforms have emerged, such as superconducting qubits~\cite{RevModPhys.93.025005}, trapped ions \cite{doi:10.1063/1.5088164} and semiconductor spin qubits~\cite{doi:10.1146/annurev-conmatphys-030212-184248}. The implementation of spin qubits in silicon~\cite{Zwanenburg_2013} and germanium~\cite{hendrickx2020fourqubit} has enabled very long coherence times and high fidelity spin manipulations, hence these systems qualify for the scale-up to large-scale qubit devices. In recent work \cite{xue2021computing} spin qubits also reveal to be a promising candidate for error-correction codes in the noisy intermediate-scale quantum (NISQ) era.
Any single-qubit gate on spin qubits in gate defined quantum dots can be generated out of single qubit rotations around two independent axes. This can be realized via electron spin resonance (ESR)~\cite{Koppens_2006} using a global stripline or through electric dipole spin resonance (EDSR) \cite{Nowack_2007} by modulating gate voltages of quantum dots within a magnetic gradient field. Together with a two-qubit operation such as the CNOT gate, this forms a universal set of qubit operations. A CNOT gate can be implemented by switching on the exchange interaction in an inhomogeneous magnetic field while driving an oscillating (effective) magnetic field on the qubits \cite{Russ_2018}. However,
unwanted interactions with the environment yielding crosstalk, dephasing and charge noise remain the fidelity-limiting factors for spin qubits. Operating the exchange interaction at a symmetric operation point (''sweet spot'') \cite{PhysRevLett.116.116801, PhysRevLett.116.110402, PhysRevLett.115.096801} suppresses charge noise to first order, while the large energy splitting due to a strong magnetic field gradient~\cite{nichol2016highfidelity} realized by a micromagnet \cite{Yoneda_2015, Kawakami_2014} reduces dephasing effects.

\begin{figure}[t]
	\centering
	\includegraphics[width=0.48\textwidth]{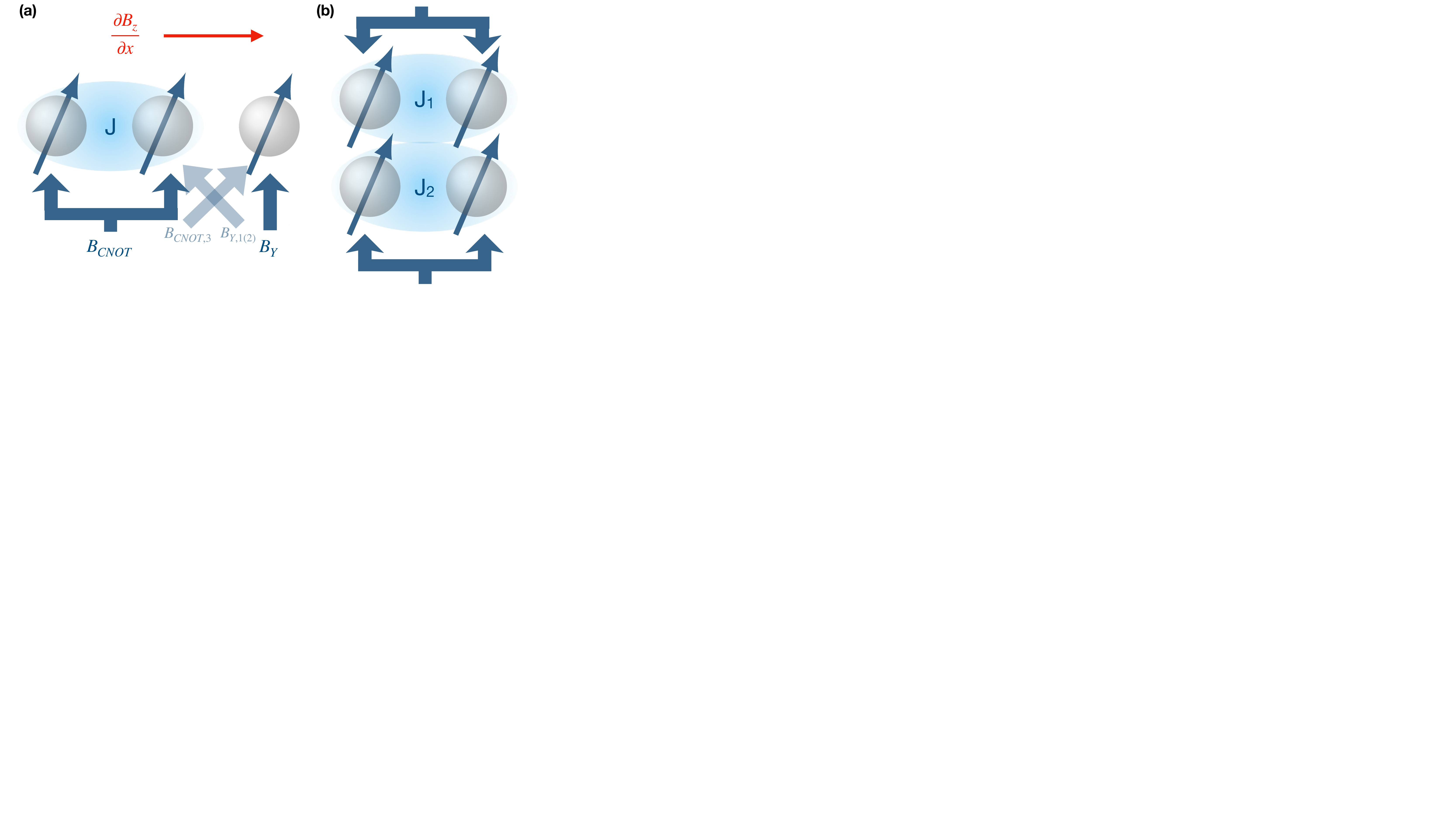}
	\caption{Quantum gates in qubit arrays with a gradient field along the $x$ axis. \textbf{(a)} Simultaneously driven CNOT and $Y$ gates with driving strengths $B_{\text{CNOT}}$ and $B_{Y}$, crosstalk fields $B_{\text{CNOT},3}$ and $B_{Y,1 (2)}$ and exchange interaction $J$. \textbf{(b)} Two parallel CNOT gates in a two-dimensional array with exchange interactions $J_1$ and $J_2$.}
	\label{Fig:setup}
\end{figure}

After high fidelity CNOT gate implementations have been proposed and demonstrated in a Si/SiGe heterostructure double quantum dot architecture \cite{Russ_2018, Zajac_2017}, we focus on the effect of crosstalk induced by a residual driving amplitude from nearby operations \cite{fedele2021simultaneous, Nature555.25766, PhysRevX.9.021011, lawrie2021simultaneous}. In previous work crosstalk was already described for protocols where single qubit operations and the CNOT were driven individually and only acted on freely evolving nearby qubits, as well as for the case of two simultaneously operated $Y$ gates~\cite{PhysRevB.104.045420}. Here we extend this analysis to driving induced crosstalk on quantum gates and give examples of protocols with simultaneously driven $Y$ and CNOT gates, and two CNOT gates arranged in a two-dimensional manner (Fig.~\ref{Fig:setup}). When scaling up systems this represents a new phenomenon that needs to be addressed. In the presence of two similar but distinct drive frequencies and non-negligible driving strengths the system cannot be treated using the rotating wave approximation (RWA), but can be approximated by applying the Floquet-Magnus expansion (FME). We disregard pure gate errors affecting operating qubits, which have been studied extensively~\cite{Yoneda_2017,PhysRevLett.116.116801,PhysRevX.9.021011}, and only focus on crosstalk effects.

This paper is organized as follows. In Section \ref{sec:model} we introduce the Heisenberg model and time-evolution approximation used for the analysis in Section \ref{sec:analysis}, where we discuss crosstalk on single-qubit and two-qubit operations, such as the $Y$, CNOT and CPHASE gates. In Section \ref{sec:analysisY} we first consider simultaneously driven $Y$ and CNOT gates, where we find a synchronization condition similar to the one found for two parallel $Y$ gates~\cite{PhysRevB.104.045420}. Finally we apply the same technique to investigate crosstalk in two simultaneously driven CNOT gates and show the advantage of same driving times for both operations in Section \ref{sec:analysisCNOT}. Finally, Sec.~\ref{sec:conclusion} contains our conclusions.

\section{Theoretical Model \label{sec:model}}
We consider a gate-defined quantum dot array which is operates in the $(1,1,\ldots)$ charge regime. However, our model can in principle be applied to arbitrary exchange-coupled spin qubit registers with external control. The exchange interaction between two spins is tuned by the middle barrier gates, as in Refs.~\cite{Russ_2018} and \cite{Zajac_2017}. For our analysis we neglect excited valley states and spin-orbit coupling, which is weak for electrons in silicon. The system is then described by the Heisenberg Hamiltonian
\begin{align}
	H = \sum_{\langle i,j\rangle} J_{ij}(t) \left( \mathbf{S}_i \cdot \mathbf{S}_j -\frac{1}{4}\right) + \sum_{i} \mathbf{S}_i \cdot \mathbf{B}_i,
\end{align}
where $J_{ij}$ is the tunable exchange interaction between two neighboring spins $\mathbf{S}_i$ and $\mathbf{S}_j$, denoted by $\langle i,j\rangle$, which is required for two-qubit operations, and $\mathbf{B}_i = (0,B_{y,i}(t),B_{z,i})$ denotes the external magnetic field at the position of spin $\mathbf{S}_i$. We provide magnetic fields in energy units throughout this paper, i.e., $\mathbf{B}_{\mathrm{physical}}=\mathbf{B}/g\mu_B$, and we furthermore set $\hbar = 1$. The magnetic field consists of a large homogeneous field plus a  gradient field (e.g. induced by a micromagnet) pointing in $z$-direction and varying along the qubit array, i.e. the $x$-axis, $B_{z,i}=B_{z} + b_{z,i}$, allowing individual addressability of single spins. A slight field gradient in $y$-direction $B_{y0,i}$ allows EDSR by applying time-dependent electrical driving, which results in a magnetic driving field in $y$-direction $B_{y,i}(t) = B_{y0,i} + B_d(t)$, with
\begin{align}
    B_d(t) = B_{d,i} \cos(\omega_d t) + B_{\text{CT},i} \cos(\omega_{\text{CT}} t + \phi),\label{acdrive}
\end{align}
where $B_{d,i}$ and $\omega_d$ denote the driving amplitude and frequency of the respective qubit operation and $B_{\text{CT},i}$ and $\omega_{\text{CT}}$ the driving amplitude and frequency of the crosstalk field originating e.g. from a close by qubit operation, respectively. We assume $\omega_{d}$ and $\omega_{\text{CT}}$ to be nearby but distinct frequencies for the case of neighboring qubit operations.
In general this can describe ESR or EDSR, where the effective magnetic driving strength for EDSR is proportional to the electric field depending on the device architecture, spin-orbit coupling mechanism, and gate voltage \cite{PhysRevB.74.165319, PhysRevLett.96.047202}. In the rotating frame $\tilde{H}(t) = R^{\dagger}HR+i\dot{R}^{\dagger}R$ with $R=\exp(-i \omega_d t \sum_{i} \mathbf{S}_i)$ we make the rotating wave approximation, neglecting fast oscillating terms. Nevertheless, time-dependent driving terms from neighboring qubit operations remain due to similar frequencies and do not cancel out in the RWA. To calculate first order corrections to the RWA time evolution, as in ideally isolated qubit operations, we take into account higher-order corrections within the Floquet-Magnus expansion for periodically driven systems (see Appendix \ref{appFM}) \cite{Bukov_2015,Blanes_2009,Moore_1990,Mostafazadeh_1997}. Crosstalk effects on single-qubit and two-qubit gates from the neighboring qubit operations can then be evaluated using the fidelity~\cite{Pedersen_2007} $F~=~(d+| \text{Tr}[ U_{\text{ideal}}^{\dagger}U_{\text{actual}} ] |^2 )/(d(d+1)) ,\label{fidelityeq}$ where $d$ is the dimension of the Hilbert space, $U_{\text{ideal}}$ is the desired qubit operation, which in our case would be the crosstalk-free qubit operation, and $U_{\text{actual}}$ the actual operation containing unwanted off-resonant Rabi-like oscillations.

\section{Crosstalk analysis}\label{sec:analysis}
We now introduce the magnetic crosstalk field $B_{\text{CT}}$ acting on a qubit operation which for EDSR can be capacitively induced \cite{Cayao_2020,PhysRevApplied_12_064049} by a driving field applied on nearby qubits \cite{Nowack_2007}. The case of ESR with a global drive is also covered by our model. To give concrete and realistic examples, in the following we consider qubit gate operations on a qubit with resonance frequency $(2\pi)\,18.493$~MHz \cite{Zajac_2017} for the single-qubit gate, and on two qubits with single-qubit frequencies $(2\pi)\,18.493$~MHz and $(2\pi)\,18.693$~MHz for a two-qubit gate. For the crosstalk frequency $\omega_{\text{CT}}$ we assume $(2\pi)\,18.693$~MHz for the single-qubit gate and $(2\pi)\,18.893$~MHz ($(2\pi)\,18.823$~MHz) for the CNOT (CPHASE) gate. 
Definitions of the studied gates are provided in Appendix \ref{App:quantumgates}.

\subsection{Crosstalk on $Y$ gates} \label{subsec:Ygate}
\begin{figure*}[t]
	\centering
	\includegraphics[width=0.98\textwidth]{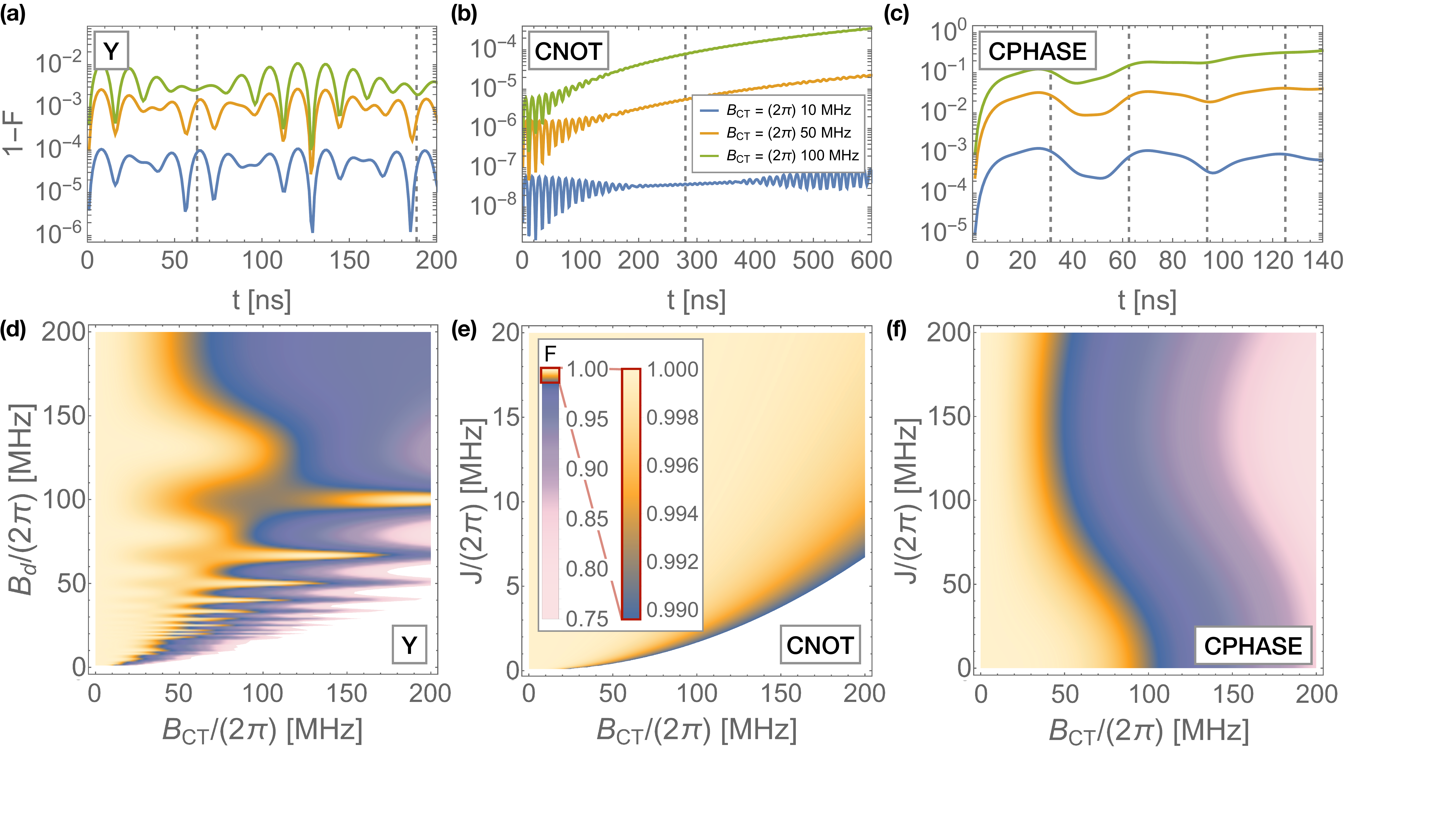}
	\caption{\textbf{(a)-(c)}: Infidelity of a \textbf{(a)} $Y$, \textbf{(b)} CNOT and \textbf{(c)} CPHASE gate suffering from crosstalk depending on the driving time. Crosstalk fields $B_{\text{CT}}$ are set to $(2\pi)\, 10$ MHz, $(2\pi)\, 50$ MHz and $(2\pi)\, 100$ MHz. Grey dashed vertical lines indicate the corresponding $Y$, CNOT and CPHASE gate times. The actual driving strength of the $Y$ and CNOT gates are $(2\pi)\,100$~MHz and $(2\pi)\,21.358$~MHz, the exchange interaction in the CNOT and CPHASE is $J = (2\pi)\,20$~MHz. \textbf{(a)}~The frequency of the oscillating behaviour in the $Y$ gate infidelity corresponds to the difference between driving and crosstalk frequencies. \textbf{(b)}~Small and fast oscillations of the CNOT gate infidelity are negligible compared to the overall fidelity impairment over time. \textbf{(c)}~For the same value of $J$, crosstalk has the largest impact on the CPHASE gate. The infidelity increases with longer driving, and the shape of the time-dependent infidelity curve strongly depends on the driving and resonance frequencies of the qubits. \textbf{(d)-(f)}:~Fidelity of the $Y$, CNOT and CPHASE gate after the corresponding driving times $\tau_{Y}$, $\tau_{\text{CNOT}}$ and $\tau_{\text{CPHASE}}$. The driving requencies are 18.693 MHz, 18.893 MHz and 18.823 MHz, respectively. \textbf{(d)}~As in (a), oscillations along the $B_{d}$ axis correspond to the difference between driving and crosstalk frequencies. \textbf{(e)}~A large exchange interaction and a small crosstalk field decrease crosstalk effects in the CNOT gate performance. \textbf{(f)}~For a CPHASE with phase $\varphi$ approaching $\pi$ the fidelity decreases.}
	\label{Compare}
\end{figure*}

Previous work \cite{PhysRevB.104.045420} investigated the impact of crosstalk on a $Y$ gate for the special case when simultaneously driving a second $Y$ gate nearby. We generalize this analysis to any crosstalk driving field. Therefore, we compare the $Y$ gate rotation at time $t$ to the actual operation at the same time for different crosstalk fields in Fig. \ref{Compare}~(a). Apparently, superposed oscillations occur, where amplitudes become large for $2\pi$ rotations and small for $\pi$ rotations.
The time evolution within the RWA is simplified when choosing $B_d$ such that the $Y$ gate driving time $\tau_Y$ fulfills $\tau_Y = 2\pi(2k+1)/B_{Y,3} = 2 \pi l/(\omega_{Y}-\omega_{\text{CNOT}})$ with $k,l \in \mathbb{Z}$, i.e., the gate time equals an integer multiple of the period of around 16 ns of the high-frequency beating oscillations in Fig.~\ref{Compare}~(a). Similar to Ref.~\cite{PhysRevB.104.045420} we find a second period in the evolution operator, which in principle can improve the operation, nevertheless, due to the high increase of driving time, a simultaneous driving seems ineffective compared to sequential operations. In Fig.~\ref{Compare}~(d) The $Y$ gate fidelity is shown as a color plot depending on the crosstalk field $B_{\text{CT}}$ and the driving strength of the $Y$ operation $B_d$. For driving amplitudes within $(2\pi)\,100$~MHz an oscillating behavior is found. The fidelity maxima correspond to driving strengths fulfilling the above condition for $\tau_Y$, which thus turn out to be advantageous. A similar behavior is also obtained for the $X$ gate.

\subsection{Crosstalk on CNOT and CPHASE gates}\label{subsec:CNOT+CPHASEgate}
A CNOT gate can be implemented by adiabatically switching on the exchange interaction $J_{12} = J$ between two neighboring qubits 1 and 2, where $J$ is much smaller than the magnetic gradient field $\Delta B_z$ in $z$ direction ($J\ll \Delta B_z$) to shift energy levels such that distinct transition frequencies allow individual addressing of the $|10\rangle\leftrightarrow |11\rangle$ transition \cite{Russ_2018, Zajac_2017}. A driving field matching this transition frequency results in a CNOT gate ($\Omega_{\text{CNOT}}$), where the off-resonant Rabi oscillation ($\Omega_{\text{off}}$) can be cancelled out by a synchronization of Rabi frequencies $\Omega_{\text{CNOT}} = (2m+1)/(2n) \tilde{\Omega}_{\text{off}}$ with $m, n\in\mathbb{Z}$ which leads to a $J$-dependent choice of $B_{\text{CNOT}}$ \cite{Russ_2018}. We assume that the middle-barrier gate determining $J$ has no effective capacitive coupling by using virtual gates \cite{Mills_2019, Hensgens_2017,Hsiao_2020}, such that the remaining crosstalk effect is a residual crosstalk driving field with frequency $\omega_{\text{CT}}$ different from the resonant CNOT driving frequency $\omega_{\text{CNOT}} = (B_{z,1} + B_{z,2} + J - \sqrt{(B_{z,2}-B_{z,1})^2+J^2})/2$. Due to finite off-diagonal terms in the Hamiltonian, the CNOT operation is limited in the fidelity for isolated qubits 1 and 2. Since this is not a crosstalk effect we neglect this fidelity decrease in our analysis and choose the ideal operation in equation \eqref{fidelityeq} to be the slightly unideal CNOT operation, which in fact is the actual time evolution that will be acted on the qubits. To understand the impact of crosstalk induced by simultaneously driving a neighboring gate we calculate the time dependent fidelity for different crosstalk strengths in Fig. \ref{Compare}~(b), where we set $B_{\text{CT}} = (2\pi)\,10$~MHz, $(2\pi)\,50$~MHz and $(2\pi)\,100$~MHz. Here we chose $B_{\text{CT}}$ to be equal on both qubits. Apparently, for all driving strengths the fidelity dominantly decreases over time while rather small high-frequent oscillations occur. It appears to be advantageous to keep the simultaneous driving time short for the CNOT operation, rather than to find a synchronization condition minimizing crosstalk as in the $Y$ gate.
To confirm this statement we plot the fidelity of the CNOT operation after the driving time $t= \tau_{\text{CNOT}} = |2\pi (2m + 1)/(B_{\text{CNOT}}(1 + J/(2\Delta B_z)))|$ depending on residual crosstalk driving strength on the qubits and exchange interaction $J$ in Fig. \ref{Compare}~(e).
It turns out that higher crosstalk fields reduce the fidelity and higher values for $J$ increase the fidelity due to shorter driving times. Since we assume $J\ll \Delta B_z$ the choice of $J$ is bound by $(2\pi)\,20$ MHz. 

To investigate crossstalk effects in a CPHASE we assume an instant $J$ pulse between control and target qubit, where $J$ is not bound to the low frequency regime. Free evolution until time $\tau_{\text{CPHASE}} = 2\pi m /\sqrt{\Delta B_z^2 + J^2}$ with $m\in \mathbb{Z}$ will then result in a CPHASE gate with phase $2\pi m J/\sqrt{\Delta B_z^2 + J^2}$. If we include a crosstalk drive in our description we obtain a time-dependent fidelity plot as in Fig. \ref{Compare}~(c). Similar to the CNOT the overall infidelity increases with progressing time, and we obtain an alternating behaviour with larger amplitude and longer period than in the CNOT gate. The shape and position of the dips strongly varies with the driving frequency, strength, and the resonance frequencies of the qubits, and thus, they are not necessarily located at the driving times. Moreover, for the same value of $(2\pi)\,20$ MHz for the exchange interaction crosstalk effects are more significant for the CPHASE than for the CNOT operation. This is also confirmed by Fig. \ref{Compare}~(f), where for the various $J$ and $B_{\text{CT}}$ the fidelity ranges from 0.8 to 1, different from the range in a similar plot for the CNOT in (b). Similar as the previous gates, CPHASE operations suffering from crosstalk operate more accurately with low crosstalk fields. Apparently, choosing $J$ such that the phase holds $\pi$, i.e. $J = (2\pi)\,115.47$ MHz, increases the impact of crosstalk and thus the infidelity.

In both considered two-qubit gates, calculations of crosstalk effects on only one of the involved qubits show that a disturbed target qubit has a larger impact on the fidelity than errors on the control qubit.

\section{Examples for parallel gates}\label{sec:examples}
In the following, we consider a CNOT gate performed on qubits 1 and 2 with driving strength $B_{\text{CNOT},1} = B_{\text{CNOT},2} = B_{\text{CNOT}}$ and a simultaneous single-qubit (CNOT) operation on qubit 3 (3 and 4) with driving strength $B_{Y,3} = B_{Y}$ ($B_{2,1} = B_{2,2} = B_{2}$). We then study their corresponding crosstalk on the respective neighboring operation in the form of an unwanted magnetic driving field on the neighboring qubit as shown in Fig.~\ref{Fig:setup}. Here we indicate magnetic ac fields on qubit $i \in \{1,2,3,(4)\}$ by $B_{\beta,i}$, where $\beta=\text{CNOT}, Y (2)$.
Furthermore, we assume the crosstalk to be symmetric, i.e. the ratio $B_{\text{CT}}/B_{d}$ between the crosstalk field from a neighbor operation and the driving strength of the actual operation coinside for both gates. Therefore we define $\alpha$ as the crosstalk parameter between qubit 2 and 3 (and 1 and 4) and $\tilde{\alpha}$ describing crosstalk between qubit 1 and 3 (and 2 and 4), such that $B_{\text{CNOT},3} = (\alpha+\tilde{\alpha}) B_{\text{CNOT}}$, $B_{Y,1} = \tilde{\alpha} B_{Y}$ and $B_{Y,2} = \alpha B_{Y}$ ($B_{\text{CNOT},3 (4)} = (\alpha+\tilde{\alpha}) B_{\text{CNOT}}$ and $B_{\text{2},1 (2)} = (\alpha+\tilde{\alpha}) B_{\text{2}}$). Residual driving fields as crosstalk between qubit 1 and 2 (or 3 and 4) are neglected, since the driving strength is the same on both qubits.

\subsection{Simultaneously driven CNOT gate and single qubit rotation}\label{sec:analysisY}
Here, we consider a CNOT gate on qubits 1 and 2 with exchange coupling $J_{12} = J$ ($J\ll \Delta B_z$). Due to the CNOT synchronization condition the CNOT driving field $B_{\text{CNOT}}$ is determined by the choice of $J$. Here, we assume resonance frequencies $B_{z,1}$, $B_{z,2}$ and $B_{z,3}$ of qubits 1, 2 and 3 to be $(2\pi)\,18.493$~MHz, $(2\pi)\,18.693$~MHz and $(2\pi)\,18.893$~MHz \cite{Zajac_2017}.
The CNOT is driven with frequency $\omega_{\text{CNOT}}$ and the $y$ rotation on qubit 3 is driven with its resonance frequency $\omega_Y = B_{z,3}$. Using $\alpha$ and $\tilde{\alpha}$ as crosstalk parameters, and defining $B_{Y,i}$ as the $Y$ gate driving on qubit $i=1,2,3$, the residual fields on the neighboring qubits fulfill $B_{Y,2} = \alpha B_{Y,3}$, $B_{Y,1} = \tilde{\alpha} B_{Y,3}$ and $B_{\text{CNOT},3} = \alpha B_{\text{CNOT},2} + \tilde{\alpha} B_{\text{CNOT},1}$. To perform a $\pi$ rotation around the $y$ axis, the phase $\phi$ is set to zero.

As the description in Section \ref{subsec:CNOT+CPHASEgate} shows the impact of crosstalk on the CNOT gate operation in Fig. \ref{Compare}~(b) and (e), the choice of a comparably large $J$ of $(2\pi)\,20$~MHz and short simultaneous driving times result in smaller errors (i.e., infidelity $1-F$).

In order to investigate the $Y$ gate fidelity suffering from crosstalk induced by the neighboring CNOT drive, we compare the $Y$ gate rotation at time $t$ to the actual operation at the same time for different crosstalk parameters $\alpha$ and $\tilde{\alpha}$ similar as in Fig. \ref{Compare}~(b). 
In contrast to the above results for the CNOT the fidelity has repeated maxima at rotation angles that are multiples of $2\pi$. In addition to the condition for the Y gate, similar to the simultaneous $Y$ gates in Ref.~\cite{PhysRevB.104.045420}, we find a periodicity of the time evolution, which in principle could be utilized to optimize the fidelity; however, the use of several full rotations results in a long driving time for the CNOT gate, and thus large infidelities in the CNOT operation. 
Instead, we choose a protocol where the CNOT and $Y$ gates are simultaneously driven for the shortest $Y$ gate time $\tau_Y$, with $k=0$, which is synchronized with the period of 
$|3 (2\pi)/(\omega_{Y}-\omega_{\text{CNOT}})|$ by choosing the drive strength appropriately (in our numerical example, $B_{Y,3} = 130.42$~MHz). Then, the remaining CNOT dynamics is performed in the absence of crosstalk, while the remaining free evolution of qubit 3 is synchronized by choosing $J$ appropriately. In our example, $J=18.32$ MHz, assuming $\alpha=0.1$ and $\tilde{\alpha} = 0.01$. These are realistic values one could find in a one dimensional qubit chain. For this set of values we find an infidelity of order $10^{-8}$ for the CNOT and $10^{-6}$ for the $Y$ gate. Fig.~\ref{CNOT-CNOT}~(a) shows the infidelities of the $Y$ and CNOT gates compared to their crosstalk-free evolution at the complete gate times $\tau_Y$ and $\tau_{\text{CNOT}}$. Since after the time $\tau_Y$ the residual crosstalk on the $Y$ gate by a single drive can be avoided by synchronization~\cite{PhysRevB.104.045420} there is no further impairment of the fidelity.

As a further test, we have introduced a non-trivial phase $\phi$ in Equation~\eqref{acdrive} in order to study its effect on the gate operation. We find only a slight  dependence of the fidelity on the phase. Since the adjustment of $\phi$ can be seen as applying an artificial $z$ rotation our observation implies a similar behavior for the $X$ gate.

As an alternative to the scheme outlined above, a complete CNOT rotation can be carried out while driving both gates simultaneously, and subsequently finishing the $Y$ gate in the absence of the CNOT drive, while fulfilling the crosstalk synchronization for the remainder of the rotation.  When doing so, one has to take into account that a finite $J$ will cause a CPHASE-like evolution in the absence of crosstalk and thus has to be tackled by applying a suitable adiabatic pulse or by adjusting the driving time for the dc pulse. Since this control sequence would lead to longer driving times, we find the above protocol more effective.

The CNOT operation is not symmetric in the sense that the target qubit performs a $\pi$ rotation while the control qubit does full $2\pi$ rotations. To investigate each possible constellation of the involved qubits, the position of qubit 3 or the position of control and target qubit within the CNOT can easily be changed by adjusting the crosstalk parameters $\alpha$ and $\tilde{\alpha}$ and modifying the magnetic field gradient in $z$ direction.

\subsection{Simultaneously driven CNOT gates} \label{sec:analysisCNOT}
\begin{figure*}[ht]
	\centering
	\includegraphics[width=0.99\textwidth]{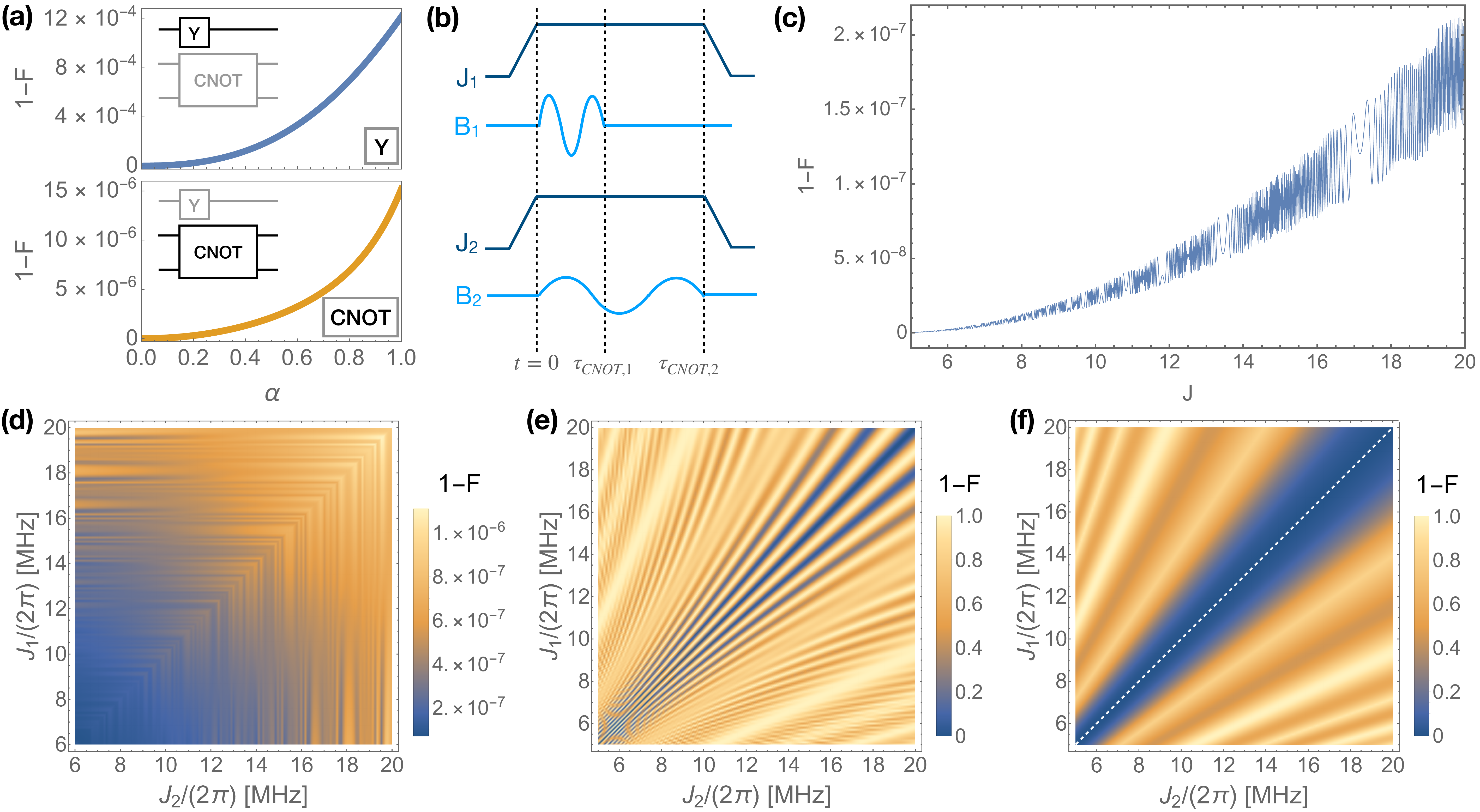}
	\caption{ \textbf{(a)} Infidelities $1-F$ of simultaneously performed $Y$ and CNOT gates depending on $\alpha$ where $\tilde{\alpha} = \alpha^2$. Upper panel: $1-F$ of the $Y$ gate after its gate time $\tau_Y$; lower panel: $1-F$ for the CNOT operation after its gate time $\tau_{\text{CNOT}}$. After $\tau_Y$ the qubit frequency of the $Y$ gate has been synchronized to completely avoid further crosstalk. \textbf{(b)} Pulse sequence of two simultaneous CNOT gates, where $J_1 \neq J_2$ leads to unequal driving strengths. \textbf{(c)} Diagonal cut of figure (f) (white dashed line): Infidelity depending on the exchange interaction $J$, where $J_1=J_2=J$. The oscillating behavior originates from crosstalk. \textbf{(d)-(f)} Color plots of the total system's infidelity after \textbf{(d)} the shorter driving time compared to the intended operation at this time, \textbf{(e)} the longer driving time and \textbf{(f)} after subsequent correcting $z$ rotations. The overall dominant pattern is due to some CPHASE-like time evolution of the qubit pair with shorter driving time.}
	\label{CNOT-CNOT}
\end{figure*}

Two-qubit gates are key ingredients of quantum algorithms as they enable entangled states. Thus, the availability of simultaneously driven CNOT gates on neighboring qubits is crucial, either in one- or two-dimensional qubit arrays. Here, we investigate the impact of crosstalk in possible constellations of two adjacent CNOT operations. Since we assume a finite exchange interaction only between the control and target qubit within one CNOT operation, we can describe each qubit pair separately. Considering one qubit pair, we can introduce crosstalk parameters $\alpha$ and $\tilde{\alpha}$ again, and already obtain the same description as in the previous case for simultaneous CNOT and $Y$ gates. Here, a short driving time turned out to be beneficial. However, instead of the $Y$ gate, we obtain the same description for the first and second CNOT gate. Hence, unlike in the case of the $Y$ gate, a synchronization condition for this case could not be found. Additionally, both driving fields are restricted by the CNOT synchronization condition and determined by the choice of $J$. The device architecture and properties are included in our description by parameters $\alpha$ and $\tilde{\alpha}$. For the special case of a two-dimensional qubit array with a fixed magnetic field gradient along the $x$ axis (see Fig.~\ref{Fig:setup}~(b)) and symmetric crosstalk we set $\alpha+\tilde{\alpha} = 1$. In this case the position of the control and target qubits is not relevant. The exchange couplings $J_1$ and $J_2$ of the first and second qubit pair represent the free parameters of the problem that determine the driving frequencies, strengths and times. For our analysis we assume that both gates are simultaneously driven with the shorter driving time, which can be described within the FME. Subsequently, the CNOT with the long driving time is finished with only a single drive on the corresponding qubit pair until the full CNOT rotation is completed, as e.g. in Fig.~\ref{CNOT-CNOT}~(b). We consider the case where the exchange interaction of the other qubit pair is still switched on, which results in a CPHASE-like time evolution suffering from crosstalk from the ongoing operation nearby. 
In Fig.~\ref{CNOT-CNOT}~(d)-(f) the infidelities of the total system, including both CNOT gates are shown for different times of the sequence. The color plots depend on the exchange interactions $J_1$ and $J_2$ of qubit pairs 1 and 2. All plots are symmetric since the fidelities of the single qubit pairs are mirror-symmetric to each other. In Fig.~\ref{CNOT-CNOT}~(d) the fidelity after the shorter driving time (corresponding to the larger exchange) shows an oscillating behavior, where oscillations within infidelities of $10^{-7}$ are rather small. After the longer CNOT time the pattern evolves to Fig.~\ref{CNOT-CNOT}~(e) featuring small and large stripe patterns. Applying subsequent $z$ rotations results in pattern Fig.~\ref{CNOT-CNOT}~(f), where only the large stripes remain. These are due to the disturbed CHPASE-like time evolution and can be avoided to a large extent by choosing equal exchange strengths $J_1$ and $J_2$. In Fig.~\ref{CNOT-CNOT}~(c) the infidelity of a cut along the diagonal of Fig.~\ref{CNOT-CNOT}~(f) is shown. Since in this special case the driving times are equal, we find that no subsequent time evolution has to be cancelled and no further crosstalk occurs. This case exhibits a behaviour that was already indicated by Fig.~\ref{CNOT-CNOT}~(d) where the infidelity increases with increasing strength of the exchange coupling. Other than in the case of the combination of CNOT and $Y$ gates, and despite the fact that the driving time becomes shorter, an increasing $J$, and thus driving field, impairs the operation quality. This effect appears due to an increasing crosstalk field on the neighboring qubit and remains for finite $\alpha$ and $\tilde{\alpha}$. Nevertheless, the  infidelity shown in Fig.~\ref{CNOT-CNOT}~(c) ranges up to the order of $10^{-7}$ for a crosstalk field as large as the adjusted driving field. This is already an extreme assumption, and thus in real applications the infidelity will probably be even smaller. 

We also considered a special case with equal resonance frequencies, which is generally not given in every setup. However, the overall effect of unavoidable crosstalk and subsequent CPHASE-like time evolution for different driving times will appear in every setup. Also, the direction of the magnetic gradient changes the driving strengths and times and thus leads to asymmetric patterns compared to \ref{CNOT-CNOT}(d)-(f). In principle, any architecture can be described this way by adjusting the crosstalk parameters $\alpha$ and $\tilde{\alpha}$ and resonance frequencies.

\section{Conclusions \label{sec:conclusion}}
In this paper we extended the analysis of crosstalk effects in spin qubit devices by investigating the fidelities of single-qubit and two-qubit operations suffering from a residual off-resonant drive. We found specific fidelity-improving conditions for the driving strengths in single-qubit gates, such as the $Y$ gate, and we showed that crosstalk in two-qubit operations, as in a CNOT and CPHASE, causes an unpreventable fidelity impairment which can mainly be reduced by choosing short driving times and a large exchange interaction in the allowed regime of $J\ll \Delta B_z$ for a given crosstalk strength. We also found the target qubit within the CNOT and CPHASE to be more sensitive to crosstalk fields than the control qubit.

We further provided examples for combinations of simultaneously driven quantum gates. In agreement with previous studies, we found the $Y$ gate to be improved by synchronizing the driving strength of the $Y$ gate itself and by fulfilling the crosstalk synchronization from Ref.~\cite{PhysRevB.104.045420}.
However, when driving two CNOT gates in parallel the synchronization of Rabi frequencies within one CNOT gate restricts the choice of driving strengths and times by the choice of the exchange interaction. A significant decrease in the fidelity originates from the CPHASE-like time evolution suffering from crosstalk when the shorter CNOT is finished but the exchange interaction is not yet fully switched off. Thus it is advantageous to avoid a residual exchange interaction or to use equal driving times by choosing exchange interactions properly.

In conclusion, our numerical results show the non-trivial behavior of single-qubit and two-qubit gates suffering from crosstalk induced by residual driving strengths. Our studies can be used to characterize systems affected by crosstalk, and so find the appropriate electric and magnetic fields for optimal control used to e.g. run quantum algorithms on spin qubits as quantum computing hardware. However, to further minimize these effects in spin qubits optimal control sequences avoiding nearest neighbour driving as dynamical decoupling and pulse shaping \cite{PhysRevB.97.045431, PhysRevLett.108.086802, PhysRevB.96.035441} and ac driven virtual gates are required.

\section*{Acknowledgments}
This work has been supported by QLSI with funding from the European Union's Horizon 2020 research and innovation programme under grant agreement No 951852 and by the Deutsche Forschungsgemeinschaft (DFG, German Research Foundation) Grant No.\ SFB 1432 - Project-ID 425217212.

\appendix

\section{Floquet-Magnus expansion}\label{appFM}
To investigate the crosstalk induced by simultaneously driving neighboring qubits we determine the corrections to the rotating wave approximation by calculating the first two orders of the Floquet-Magnus expansion for the time-evolution 
\begin{align}
U(t)=\text{T}\exp\left(-i\int_0^t H(\tau)\text{d}\tau\right)
\end{align}
generated by a periodic Hamiltonian $H(t)=H(t+T)$ with period $T$ \cite{Bukov_2015,Blanes_2009,Moore_1990,Mostafazadeh_1997}.
For this purpose, one defines
\begin{align}
    &F_1 = \frac{1}{T} \int_0^T (-i H(\tau)) \text{d}\tau,\\
    &\Lambda_1 (t) = \int_0^t (-i H(\tau)) \text{d}\tau - t F_1 ,\\
    &F_2 = \frac{1}{2T} \int_0^T [-i H(\tau)+F_1,\Lambda_1(\tau)] \text{d}\tau ,\\
    &\Lambda_2 (t) = \frac{1}{2} \int_0^t [-i H(\tau) + F_1, \Lambda_1(\tau)] \text{d}\tau - t F_2 .
\end{align}
Time evolution is then approximately given by
\begin{align}
    U(t) \approx \exp\left(\sum_{i=1}^2 \Lambda_i (t)\right)\exp\left(\sum_{i=1}^2 F_i t\right)
\end{align}
and can be used to describe two simultaneous ac pulses with different frequencies. Here we first go into the rotating frame of the main driving frequency $\omega_d$, where we apply the RWA and use the FME to correct terms rotating with $\omega_d-\omega_{\text{CT}}$ which are not cancelled by the RWA.  Here, $\omega_{\text{CT}}$ denotes the driving frequency of the crosstalk field originating from nearby qubit operations. Higher order corrections of magnitude  $\mathcal{O}((\omega_d-\omega_{\text{CT}})^{-2})$~\cite{doi:10.1063/1.3610943, Bukov_2015} are truncated in our analysis. With approaching frequencies $\omega_d$ and $\omega_{\text{CT}}$ higher orders must be taken into account.

\section{Quantum gates} \label{App:quantumgates}
The single-qubit gates $X$, $Y$ and $Z$ correspond to the Pauli matrices $\sigma_x$, $\sigma_y$ and $\sigma_z$. The CNOT gate in our setup is defined as the unitary matrix in the basis
$\{|00\rangle,|01\rangle,|10\rangle,|11\rangle\}$,
\begin{align}
    \text{CNOT} = \begin{pmatrix}
    1&0&0&0\\
    0&1&0&0\\
    0&0&0&1\\
    0&0&1&0
    \end{pmatrix}
\end{align}
where the first qubit act as the control qubit and performs a spin flip in case that the second qubit, the target qubit, is in state spin up.
The CPHASE gate
\begin{align}
    \text{CPHASE} = \begin{pmatrix}
    1&0&0&0\\
    0&1&0&0\\
    0&0&1&0\\
    0&0&0&e^{i\varphi}
    \end{pmatrix}
\end{align}
with phase $\varphi = 2\pi m J/\sqrt{\Delta B_z^2 + J^2}$ is accomplished by switching on the exchange interaction for time $\tau_{\text{CPHASE}} = 2\pi m / \sqrt{\Delta B_z^2 + J^2}$, where $m\in \mathbb{Z}$, plus subsequent z rotations on both qubits.

\bibliography{bibliography}

\end{document}